\documentclass{ws-rv9x6}
\usepackage{ws-rv-van}             
\makeindex
\begin{document}

\chapter{Nonequilibrium dynamical mean-field theory
of strongly correlated electrons\label{fk_model}}

\author[V.~Turkowski]{V.~Turkowski}
\address{Department of Physics and Astronomy,
University of Missouri-Columbia, Columbia, MO 65202, USA\\
turkowskiv@missouri.edu}
\author[V.~Turkowski and J.K.~Freericks]{J.K.~Freericks}

\address{Department of Physics, Georgetown University,\\
Washington, D.C. 20057, USA\\
freericks@physics.georgetown.edu
}
\begin{abstract}
We present a review of our recent work in extending the successful
dynamical mean-field theory from the equilibrium case to
nonequilibrium cases.  In particular, we focus on the problem
of turning on a spatially uniform, but possibly time varying,
electric field (neglecting all magnetic field effects).  We show how to
work with a manifestly gauge-invariant formalism, and
compare numerical calculations from a transient-response
formalism to different types of approximate treatments,
including the semiclassical
Boltzmann equation and perturbation theory in the interaction.
In this review, we solve the nonequilibrium problem for the Falicov-Kimball
model, which is the simplest many-body model and the easiest problem
to illustrate the nonequilibrium
behavior in both diffusive metals and Mott insulators. Due to space
restrictions, we assume the reader already has some familiarity both
with the Kadanoff-Baym-Keldysh
nonequilibrium formalism and with equilibrium dynamical mean-field theory;
we provide a guide to the literature where additional details
can be found.
\end{abstract}

\body

\section{Introduction}
\label{Introduction}

Dynamical mean-field theory (DMFT) was introduced in
1989~\cite{brandt_mielsch_1989} shortly after Metzner and
Vollhardt~\cite{metzner_vollhardt_1989}
proposed scaling the hopping matrix element as the inverse square
root of the spatial dimension to achieve a nontrivial limit where
the many-body dynamics are local. Since then, the field has
blossomed to the point where nearly all model many-body problems
have now been solved\cite{georges_kotliar_krauth_rozenberg_1996},
and much recent
work has focused on applying DMFT principles to real materials
calculations\cite{kotliar_savrasov_haule_2006}.
Little work has emphasized
nonequilibrium aspects of the many-body problem, where the
strongly correlated system is driven by an external field
that can possibly sustain a nonequilibrium steady state. In this
contribution, we will review recent work that has been completed
on expanding DMFT approaches into the nonequilibrium realm.
We will show how to work with so-called gauge-invariant Green
functions\cite{bertoncini_jauho} to illustrate that
one can carry out calculations in a form that manifestly is
independent of the gauge chosen to describe the driving fields.
This approach is different from our previously published work,
where we worked solely with Green functions in the Hamiltonian
gauge (where the scalar potential vanishes).

We examine the problem of strongly correlated electrons driven by a
spatially uniform electric field in the limit of infinite
dimensions\cite{Nashville,Turkowski1,dmft_fk}, where
DMFT can be applied to solve the problem exactly.
In infinite dimensions, the self-energy of the electrons
is local, and the lattice problem can be mapped onto the
problem of an impurity coupled to an effective time-dependent field
(which is adjusted so that the impurity Green function and the
local Green function on the lattice are identical).
The impurity problem in the dynamical mean field
can be solved exactly for many different
cases. In equilibrium, a large number of
strongly correlated models
have been solved in infinite dimensions, like the
Falicov-Kimball
model\cite{brandt_mielsch_1989,brandt_mielsch_1990,brandt_mielsch_1991},
the Hubbard model\cite{jarrell_1992,rozenberg_zhang_kotliar_1992,%
bulla_1999},
the periodic Anderson model\cite{jarrell_pam,%
pruschke_bulla_jarrell_2000}, and the Holstein
model\cite{freericks_jarrell_scalapino_1993,meyer_hewson_bulla_2002},
(for a reviews, see
Refs.~\refcite{georges_kotliar_krauth_rozenberg_1996} and
\refcite{freericks_zlatic_2003b}). Recently, there has been a
significant effort in
combining DMFT with density functional theory (DFT) to
describe properties of real materials when DFT is insufficient
to properly describe the electron-electron interactions
(see Ref.~\refcite{kotliar_savrasov_haule_2006}
for a review). It is now generally believed that DMFT is a good
approximation to the many-body problem in three dimensions, and it can
accurately describe strong electron-electron correlation effects in bulk
systems.

\begin{figure}
\centerline{\psfig{file=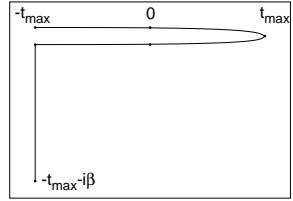,width=2.8cm, angle=270}}\caption{ The
complex Kadanoff-Baym-Keldysh time contour for the two-time Green
functions in the nonequilibrium case. The time increases from
$-t_{max}$ (left point on the top branch) along the contour to
$t_{max}$ then decreases back to $-t_{max}$ and then runs parallel
to the imaginary axis to $-t_{max}-i\beta$. We consider the
situation when the electric field is turned on at $t=0$, so the
vector potential is nonzero for $t>0$. We assume that both $t_1$ and
$t_2$ lie somewhere on the contour. } \label{fig: keldysh}
\end{figure}

The first attempt to employ DMFT to describe nonequilibrium
properties of a strongly correlated model was made by Schmidt and
Monien in Ref.~\refcite{SchmidtMonien}, where they studied the
spectral properties of the Hubbard model in the presence of a
time-dependent chemical potential by using iterated perturbation
theory (PT).  Recently, we have developed a generalized
nonequilibrium DMFT formalism to study the response of correlated
electrons to a spatially uniform time-dependent electric field and
applied that formalism to the Falicov-Kimball
model\cite{Nashville,Turkowski1,dmft_fk}. The Falicov-Kimball
model\cite{FalicovKimball}, is the simplest model for strongly
correlated electrons that demonstrates long range order and
undergoes a metal-to-Mott-insulator transition. It consists of two
kinds of electrons: conducting $c$-electrons and localized
$f$-electrons, which interact through an on-site Coulomb repulsion.
The model was introduced to describe valence-change and
metal-insulator transitions\cite{FalicovKimball} in rare-earth and
transition-metal compounds. It was reinvented as a model to describe
crystal formation\cite{KennedyLieb} resulting from the Pauli
exclusion principle. DMFT was actually developed with the original
solution of the Falicov-Kimball model in infinite
dimensions\cite{brandt_mielsch_1989,brandt_mielsch_1990,%
brandt_mielsch_1991} and now there is an
almost complete understanding of its general properties (for a
review, see Ref.~\refcite{freericks_zlatic_2003b}).
We extended the equilibrium
formalism to the nonequilibrium case, where we numerically solved a
system of the equations for the Green function and self-energy
defined on a complex time contour (see Fig.~\ref{fig: keldysh}) by
using the Kadanoff-Baym-Keldysh nonequilibrium Green function
formalism\cite{kadanoff_baym,keldysh}.

In this review, we summarize the successes of recent work to
generalize DMFT to nonequilibrium problems with a focus on solutions
of the spinless
Falicov-Kimball model on an infinite-dimensional hypercubic lattice
in the presence of an external time-dependent electric field. There
are many interesting and surprising results which differ
from semiclassical predictions (such as those made from
the Boltzmann equation
solution). In addition to the exact solutions, we also present results
for the noninteracting case and for the case of second-order
perturbation theory in the interaction. In particular, we analyze
the limitations of the perturbation theory approximation, especially
in studying (long-time) steady-state behavior.


\section{General nonequilibrium formalism}
\label{Formalism}

The nonequilibrium properties of a quantum many-particle system can
be studied by calculating the contour-ordered Green function
in momentum space:
\begin{eqnarray}
&~&G_{{\bf k}}^{c}(t_{1},t_{2})= -i \langle {\rm {\hat
T}}_{c}c_{{\bf k}H}^{}(t_{1})
c_{{\bf k}H}^{\dagger}(t_{2})\rangle
\nonumber \\
&=&\frac{-i{\rm Tr}\left\{e^{-\beta H(-t_{\rm max})} {\rm {\hat
T}_{c}}\exp [-i\int_{c}dt H_I(t)] c_{{\bf k}I}^{}(t_{1})c_{{\bf
k}I}^{\dagger}(t_{2})\right\}} {{\rm Tr}e^{-\beta H(-t_{\rm max})}},
\label{GT}
\end{eqnarray}
defined on the complex time-contour presented  in Fig.~\ref{fig:
keldysh} (see, for example, Ref.~\refcite{Rammer}).
Since the system is initially in equilibrium, the thermal
average in Eq.~(\ref{GT}) is performed with the equilibrium
density matrix $\exp[-\beta H(-t_{\rm max})]/{\rm Tr}\exp[-\beta
H(-t_{\rm max})]$ with respect to the initial Hamiltonian $H(-t_{\rm
max})$ with vanishing electric field (the symbol $\beta=1/T$ is the inverse
temperature). The operator indices $H$ and $I$ in
Eq.~(\ref{GT}) stand for the Heisenberg and Interaction
representations, respectively. In this formalism, familiar
quantum many-body techniques derived in equilibrium can also
be used in the nonequilibrium case, except that now the time
ordering ${\rm {\hat T}_{c}}$ of the operators is along the complex
contour. In particular, the Schwinger-Dyson equation, which connects
the contour-ordered Green function with the electron self-energy
$\Sigma_{{\bf k}}^c(t_{1},t_{2})$, remains valid:
\begin{equation}
G_{{\bf k}}^{c}(t_{1},t_{2}) =G_{{\bf k}}^{0c}(t_{1},t_{2}) +\int_c dt
\int_c d\bar t
G_{{\bf k}}^{0c}(t_1,t)\Sigma_{{\bf k}}^{c}(t,\bar t)G_{{\bf k}}^{c}(\bar t,t_2)
, \label{Dyson}
\end{equation}
where the matrix product of the continuous matrix operators is
accomplished by line integrals over the contour.

In DMFT, we work with the local Green function, which is found by
summing the momentum-dependent Green function over all momenta.
We then map the many-body problem on the lattice to an impurity
problem, but in a dynamical mean field that mimics the hopping
of electrons onto and off of the given site. It turns out that
one needs the full freedom available with the three-branch contour
to find the proper dynamical mean field to map the impurity onto the
lattice.  Hence, our approach will work with the less common
Green functions on the three-branch contour, as opposed to a simpler
two-branch contour, which we work with when we discuss the perturbative
approach on the lattice.  One can find the time-ordered,
anti-time-ordered, lesser, greater, retarded, advanced
and thermal Green functions on this contour\cite{wagner_1991}. For example,
the retarded Green function, which is related to the density of
quantum states, is
\begin{equation}
G^R_{\bf k}(t_1,t_2)=-i\theta(t_1-t_2)\langle \{c^{}_{{\bf k}H}(t_1),
c^\dagger_{{\bf k}H}(t_2)\}_+\rangle,
\label{eq: g_ret}
\end{equation}
where the braces indicate the anticommutator of the two operators,
and the lesser Green function, which is related to how the electrons
are distributed amongst the quantum states, satisfies
\begin{equation}
G^<_{\bf k}(t_1,t_2)=i\langle c^{\dagger}_{{\bf k}H}(t_2)
c^{}_{{\bf k}H}(t_1)\rangle.
\label{eq: g_lesser}
\end{equation}
Both of these functions can be extracted from $G^c_{\bf k}$.

\section{Nonequilibrium dynamical mean-field theory for the Falicov-Kimball model}
\label{FalicovKimball}

The spinless Falicov-Kimball model\cite{FalicovKimball}
consists of two kinds of
electrons: conduction $c$-electrons and localized $f$-electrons.
They interact with each other through an on-site Coulomb repulsion
$U$. The model Hamiltonian has the following form in the absence of
any external fields:
\begin{equation}
{\cal H}=-\sum_{\langle ij\rangle}t_{ij}c_{i}^{\dagger}c^{}_{j}
-\mu\sum_{i}c_{i}^{\dagger}c^{}_{i}
-\mu_{f}\sum_{i}f_{i}^{\dagger}f^{}_{i}
+U\sum_{i}f_{i}^{\dagger}f^{}_{i}c_{i}^{\dagger}c^{}_{i}, \label{H}
\end{equation}
where $t_{ij}=t$ is the nearest-neighbor hopping matrix element for
the $c$-electrons, $\mu$ and $\mu_{f}$ are the chemical potentials
of $c$- and $f$-electrons, correspondingly. Due to the Pauli
principle, there is no local $cc$- and $ff$-electron interaction in
the spinless case. The Hamiltonian in Eq.~(\ref{H}) can also be
regarded as an approximation to the spin $s=1/2$ Hubbard model,
where spin-up ($c$-) electrons move in a frozen background of the
localized spin-down ($f$-electrons). We consider the problem on the
infinite-dimensional ($d\rightarrow\infty$) hypercubic lattice at
half-filling, when the particle densities of the $c$- and
$f$-electrons are equal to $0.5$. In this limit, the hopping
parameter is renormalized in the following
way\cite{metzner_vollhardt_1989}: $t=t^*/2\sqrt{d}$. In the limit of
infinite dimensions, one can solve the equilibrium problem for the
conduction electrons exactly at any temperature, particle
concentration and Coulomb repulsion. The key simplification, which
allows one to obtain the exact solution as $d\rightarrow\infty$,
comes from the fact that the electron self-energy is
momentum-independent\cite{brandt_mielsch_1989,metzner}. Although
that original work was performed in equilibrium, Langreth's
rules\cite{Langreth} guarantee that it also holds for the
nonequilibrium case.

Nowadays, most of the equilibrium properties of the model, including
the phase diagram, are well known (see
Ref.~\refcite{freericks_zlatic_2003b}). In particular, the model
demonstrates a Mott transition when
$n_{c}+n_{f}=1$ at some critical value  of the Coulomb
repulsion\cite{Demchenko},
which depends on the particular value of $n_{c}$ ($n_{f}$ is equal
to $1-n_{c}$ in this case). In the insulating phase, the density of
states $A(\omega )$ is not equal to zero for frequencies inside the
``gap region'', but is exponentially suppressed, except for
$\omega =0$. Therefore, the density of states actually demonstrates a
pseudogap in the insulating phase, which is an artifact of the fact
that the infinite-dimensional hypercubic lattice has a Gaussian
density of states for the noninteracting problem, which does not have
a finite bandwidth. Another important feature is the
behavior of the imaginary part of the self-energy for frequencies
close to zero in the "metallic" phase: ${\rm Im}\Sigma (\omega )\sim -c+c^\prime
\omega^2$ ($c$ and $c^\prime >0$ and independent of temperature),
which differs from the standard Fermi liquid
behavior ${\rm Im}\Sigma (\omega )\sim -a(T)-b\omega^{2}$ ($a$ and $b>0$ and
$a(T)\rightarrow 0$ as $T\rightarrow 0$). This means that there
are no long-lived Fermi liquid quasiparticles in the model.

We are interested in the case when the system is coupled to an
external electromagnetic field ${\bf E}({\bf r},t)$. This field can
be expressed by a scalar potential $\varphi ({\bf r},t)$ and by a
vector potential ${\bf A}({\bf r},t)$ in the following way:
\begin{equation}
{\bf E}({\bf r}, t)=-{\bf \nabla}\varphi ({\bf r}, t)
-\frac{1}{c}\frac{\partial {\bf A}({\bf r}, t)} {\partial t}.
\label{Electricfield}
\end{equation}
We assume that the electric field is spatially uniform and choose
the temporal or Hamiltonian gauge for the electric field: $\varphi
({\bf r}, t)=0$. In this case, the electric field is introduced into
the Hamiltonian by means of the Peierls substitution for the hopping
matrix\cite{Jauho}:
\begin{equation}
t_{ij}\rightarrow t_{ij}\exp\left[ -\frac{ie}{\hbar c}\int_{{\bf
R}_{i}}^{{\bf R}_{j}}{\bf A}({\bf r}, t)d{\bf r} \right]
=t_{ij}\exp\left[ \frac{ie}{\hbar c} {\bf A}(t)\cdot ({\bf
R}_{i}-{\bf R}_{j}). \right] , \label{Peierls1}
\end{equation}
where the last formula holds for a spatially uniform field where we
take ${\bf A}(t)=-{\bf E}ct\theta(t)$ for a uniform field turned on at $t=0$.

We assume that it is safe to neglect magnetic field effects, because
the electric field varies slow enough in time (recall Maxwell's equations
say that a time-varying electric field creates a time varying magnetic
field). This approximation is valid
when the electric field is smooth enough in time that the magnetic fields
can be ignored.  Another way of describing this is that we assume
our electric field is always spatially uniform, even though it has
a time dependence, which is not precisely a solution of Maxwell's
equations, but is approximately so.

The electric field introduced into the Hamiltonian Eq.~(\ref{H})
results in a time-dependent shift of the momentum in the free
electron dispersion relation:
\begin{equation}
\epsilon\left({\bf k}-\frac{e{\bf A}(t)}{\hbar c}\right)
=-2t\sum_{l=1}^{d}\cos\left[a\left( k^{l}-\frac{eA^{l}(t)}{\hbar c}
\right)\right] . \label{Ek}
\end{equation}
It is convenient to consider the case, when the electric field lies
along the elementary cell diagonal\cite{Turkowski}:
\begin{equation}
{\bf A}(t)=A(t)(1,1,...,1). \label{A}
\end{equation}
In this case, the free electron spectrum
\begin{equation}
\epsilon \left({\bf k}-\frac{e{\bf A}(t)}{\hbar c}\right)
=\cos\left(\frac{eaA(t)}{\hbar c}\right)\epsilon ({\bf k})
+\sin\left(\frac{eaA(t)}{\hbar c} \right){\bar \varepsilon} ({\bf
k}), \label{energy}
\end{equation}
depends on only two energy functions:
\begin{equation}
\epsilon ({\bf k})=-2t\sum_{l}\cos (ak^{l}) \label{eps}
\end{equation}
and
\begin{equation}
{\bar \varepsilon} ({\bf k})=-2t\sum_{l}\sin (ak^{l}).
\label{bareps}
\end{equation}
Of course, when the field vanishes, the energy spectra in
Eq.~(\ref{energy}) reduces to the standard spectra in Eq.~(\ref{eps}) for
free electrons on the hypercubic lattice.
In the limit of an infinite dimensional hypercubic lattice, one can
calculate the joint density of states for the two energy functions
in Eqs.~(\ref{eps}) and (\ref{bareps})\cite{Schmidt},
\begin{equation}
\rho_{2}(\epsilon , {\bar \varepsilon}) =\frac{1}{\pi
t^{*2}a^{d}}\exp\left[ -\frac{\epsilon^{2}}{t^{*2}}-\frac{{\bar
\varepsilon}^{2}}{t^{*2}} \right] . \label{rho2}
\end{equation}
Below we use atomic units, putting all fundamental constants, except the
electron charge $e$, to be equal to one: $a=\hbar =c=t^{*}=1$.

To solve the problem of the response of the conduction electrons to
an external electric field, we use a generalized
nonequilibrium DMFT formalism\cite{dmft_fk}. The
electron Green functions and self-energies are functions of two time
arguments defined on the complex time-contour in Fig.~\ref{fig: keldysh}.
Since the action for the Falicov-Kimball model is quadratic in the
conduction electrons, the Feynman path integral over the
Kadanoff-Baym-Keldysh contour can be expressed by the determinant of a
continuous matrix operator with arguments defined on the contour.
Because the concentration of localized particles on each site
is conserved, one can calculate the trace
over the fermionic variables. It is possible to show that the
self-energy remains local in the limit of infinite dimensions in the
presence of a field; start with the
equilibrium perturbation theory expansion for the self-energy\cite{metzner}
and then apply
Langreth's rules\cite{Langreth} to the self-energy diagrams,
which say that every nonequilibrium diagram is
obtained from a corresponding equilibrium diagram, with the
time variables now defined on the Kadanoff-Baym-Keldysh contour.

The generalized system of nonequilibrium DMFT equations for the
contour ordered Green function $G (t_{1},t_{2})$, self-energy
 $\Sigma (t_{1},t_{2})$ and an effective dynamical mean-field
$\lambda (t_{1},t_{2})$ can be written in analogy with the
equilibrium
case\cite{jarrell_1992} as follows:
\begin{eqnarray}
G(t_{1},t_{2})&=&\sum_{{\bf k}}[G_{{\bf k}}^{(0)-1}-
\Sigma]^{-1}(t_{1}, t_{2}) ,
 \label{DMFT1}
\\
G_0(t_1,t_2)&=&[G^{-1}+\Sigma]^{-1}(t_1,t_2),
\label{g0_def}
 \\
\lambda (t_{1},t_{2}) &=&G^{-1}_{0imp}(t_{1},t_{2};\mu)-G_0^{-1}(t_{1},t_{2}),
 \label{DMFT2}
\\
 G(t_{1},t_{2})
&=&(1-w_{1} )G_{0}(t_{1}, t_{2}) +
w_{1}[G_{0imp}^{-1}(\mu-U)-\lambda]^{-1}(t_{1}, t_{2}),
 \label{DMFT3}
\end{eqnarray}
where $G_{{\bf k}}^{(0)}(t_{1}, t_{2})$ is the noninteracting
electron Green function in the presence of an external
time-dependent electric field, which can be calculated analytically
(see below) and $G_{0imp}(t_{1}, t_{2};\mu)$
is the free impurity Green function in a chemical potential $\mu$.
The symbol $w_{1}$ is the average number of the $f$-electrons per
site. In our case, $w_{1}=1/2$.

The momentum
summation in Eq.~(\ref{DMFT1}) can be performed by introducing the two
energy functions Eqs.~(\ref{eps}) and (\ref{bareps}) and using
the joint density of states in Eq.~(\ref{rho2}): $\sum_{{\bf
k}}F_{{\bf k}}=\int d\epsilon \int d{\bar \varepsilon}
\rho_{2}(\epsilon , {\bar \varepsilon})F_{\epsilon, {\bar
\varepsilon}}$ whenever the summand $F_{\bf k}$ depends on momentum
only through the two energy
functions. The system of equations (\ref{DMFT1})-(\ref{DMFT3})
formally resembles the corresponding system in the equilibrium case,
except now we have to work with a two-time formalism on the contour,
rather than being able to Fourier transform the relative time
to a frequency. And, because we are working with the contour-ordered
Green functions, which depend on the distribution of electrons,
we need to be careful to treat how the chemical potential is shifted by
$U$ when we perform the trace over the $f$-electrons.

The system of equations (\ref{DMFT1})-(\ref{DMFT3}) can be solved by
iteration as follows. One starts with an initial self-energy matrix, for
example the equilibrium self-energy. Substitution of this function
into Eq.~(\ref{DMFT1}) gives the Green function.
Then, from Eq.~(\ref{DMFT2}) one can find the effective
dynamical mean-field $\lambda (t_{1}, t_{2})$, which allows one to
find the new value for the Green function $G (t_{1}, t_{2})$ from
Eq.~(\ref{DMFT3}). After that, one finds the new self-energy $\Sigma
(t_{1}, t_{2})$ from the impurity Dyson equation and the dynamical
mean field. The calculations are repeated until the
difference between the old and new values for the
self-energy $\Sigma (t_{1}, t_{2})$ are smaller than some desired
precision (usually $10^{-6}$ in relative error).

In practice, to solve this system numerically, one needs to
discretize the complex time contour Fig.~(\ref{fig: keldysh}) with
some step $\Delta t$ along the real axis and $\Delta \tau$ along
the imaginary axis. In this case, the functions in
Eqs.~(\ref{DMFT1})-(\ref{DMFT3}) become general complex square
matrices. In order to
study the long time behavior, one needs to choose the value of $t_{max}$
large enough. The precision of the solution strongly depends on the
value of the discretization step $\Delta t$, which must be small
enough. Therefore, in order to get a precise long time solution it
is necessary to use large complex square matrices in
Eqs.~(\ref{DMFT1})-(\ref{DMFT3}). This causes some constraints
connected with the machine memory and the computational time.
 In our calculations, we used the time step $\Delta
t$ ranging from 0.1 to 0.0167 and matrices up to order
$4900\times 4900$.
The two-energy integration in
Eq.~(\ref{DMFT1}) was performed by using a Gaussian integration
scheme (for details, see Ref.~\refcite{Nashville}). Since each
energy is independent of each other, the algorithm parallelizes
naturally.

It is important to find ways to benchmark this nonequilibrium
DMFT algorithm, to ensure that it is accurate. The simplest way
to do this is to
calculate the equilibrium results within the nonequilibrium
formalism and compare those results with the results obtained by the
equilibrium DMFT approach. One of the most important
elements is a proper choice of the discretization step $\Delta t$ of
the contour. These equilibrium calculations can always help to
choose the step $\Delta t$ small enough to get correct results (see
Ref.~\refcite{Nashville}). Another useful way to check the
accuracy of the solution is to calculate the moments of the
electron spectral functions $A(t_{ave},\omega)=\sum_{{\bf
k}}(-1/\pi){\rm Im}G_{{\bf k}}(t_{ave}, \omega )$, where $t_{ave}$ is the
average time and
$\omega$ is the electron frequency arising from a Fourier transform of
the relative time (see below). We
have found\cite{Turkowski1} that the lowest
spectral moments in the Falicov-Kimball model
can be calculated exactly, and they are time-independent
even in the presence of a time-dependent electric field. In
particular, when a spatially homogeneous time-dependent
electric field is applied, one can find for the zeroth and first two
retarded spectral moments:
\begin{equation}
\int_{-\infty}^{\infty}d\omega A^{R}(t_{ave},\omega)=1 ,
\label{mu0Rdensity}
\end{equation}
\begin{equation}
\int_{-\infty}^{\infty}d\omega\omega
A^{R}(t_{ave},\omega)=-\mu+Un_{f}=0 , \label{mu1Rdensity}
\end{equation}
\begin{equation}
\int_{-\infty}^{\infty}d\omega
\omega^{2}A^{R}(t_{ave},\omega)=\frac{1}{2}+\mu^{2}-2U\mu
n_{f}+U^{2}n_{f} =\frac{1}{2}+\frac{U^2}{4}, \label{mu2Rdensity}
\end{equation}
where the second equality holds in the half-filled case.
We estimate the accuracy of the discretization of the contour by
calculating the spectral moments and comparing them with the exact
analytical results in
Eqs.~(\ref{mu0Rdensity})-(\ref{mu2Rdensity})\cite{Turkowski1}.
In general, one needs to reduce the discretization
size as the interaction strength increases.  This is clearly
seen in the equilibrium case, where the numerics can be well controlled
because there is no dependence on the energy $\bar\varepsilon$.
Surprisingly, in the presence of a field, one can use a somewhat
larger discretization size, especially for moderate to large fields.

\section{Gauge invariance and physical observables}
\label{Gauge}

In nonequilibrium problems, we work with two-time Green functions
because the system no longer has time-translation invariance.
Wigner\cite{wigner} first realized that it is more physical to express
results in terms of average and relative coordinates, where
the dependence on the average coordinates drops out in equilibrium.
In our case, the relative and average times satisfy
\begin{equation}
t=t_{1}-t_{2}, \ \ \ \ t_{ave}=\frac{t_{1}+t_{2}}{2}
\label{tWigner}
\end{equation}
while for the spatial coordinates we have
\begin{equation}
{\bf r}={\bf r}_{1}-{\bf r}_{2}, \ \ \ \ {\bf r}_{ave}=\frac{{\bf
r}_{1}+{\bf r}_{2}}{2}; \label{rWigner}
\end{equation}
note that at this point we are restricting the time coordinates to
lie on the real axis piece of the contour since the imaginary axis
piece is not important for determining physical properties on the lattice
(the full structure is only needed for the self-consistent
DMFT loop, not for calculating any physical properties once the
self-energy has been determined).
We want to be able to convert the relative time and space coordinates
into frequency and momentum via a Fourier transformation. Since we
are working with a uniform electric field, we expect that the
system will have no average spatial coordinate dependence, because
it is spatially homogeneous.  The easiest way to construct the right
transformation is to
create a Fourier transformation that makes the gauge-invariance
of the problem manifest; the result is then called the gauge-invariant
Green function, which depends only on the fields, not on the scalar
or vector potentials\cite{bertoncini_jauho}.
The procedure is somewhat technical, but
completely straightforward. The starting point is a generalized
Fourier transformation
\begin{eqnarray}
G({\bf k},\omega,{\bf r}_{ave},t_{ave})&=&\int d^{d} r\int dt \exp [iW ({\bf
k},\omega,{\bf r},t,{\bf r}_{ave},t_{ave})] \nonumber\\
&\times&G({\bf r},t,{\bf r}_{ave},t_{ave}),
\label{Gtransform}
\end{eqnarray}
with $W$ being a complicated function of its variables, in general.
In equilibrium, when there is no external space- and time-dependent
electric field, the Green function doesn't depend on the average
coordinates $t_{ave}$ and ${\bf r}_{ave}$, and the transform
(\ref{Gtransform}) is the well-known Fourier transformation with $W
({\bf k},\omega,{\bf r},t,{\bf r}_{ave},t_{ave})=t\omega-{\bf
r}\cdot{\bf k}$.

The situation is more complicated when an external
field is present. In this case, the field is
introduced by using a specific gauge for the scalar and vector
potential (we work with the Hamiltonian gauge).
It is important to have a Green function on the left hand
side of Eq.~(\ref{Gtransform}), which doesn't depend on the choice
of gauge so all results are manifestly independent of the scalar
and vector potentials.
Therefore, we need to construct a function
$W ({\bf k},\omega,{\bf r},t,{\bf r}_{ave},t_{ave})$ in
Eq.~(\ref{Gtransform}), which makes $G({\bf k},\omega,{\bf
r}_{ave},t_{ave})$ invariant under the gauge transformation:
\begin{equation}
\varphi ({\bf r}_1,t_1)\rightarrow \varphi ({\bf r}_1,t_1)-\frac{\partial
\chi ({\bf r}_1,t_1)}{\partial t_1},
\label{phigauge}
\end{equation}
\begin{equation}
{\bf A}({\bf r}_1,t_1)\rightarrow {\bf A}({\bf r}_1,t_1)+{\bf \nabla}\chi
({\bf r}_1,t_1), \label{Agauge}
\end{equation}
where $\chi ({\bf r}_1,t_1)$ is an arbitrary function. The $\chi$ function
must also be used in the local unitary gauge transformation of the fermion
operators:
\begin{eqnarray}
c ({\bf r}_1,t_1)&\rightarrow& \exp [ie \chi ({\bf r}_1,t_1)]c
({\bf r}_1,t_1), \\
c^{\dagger}({\bf r}_2,t_2)&\rightarrow& \exp [-ie \chi ({\bf
r}_2,t_2)]c^{\dagger}({\bf r}_2,t_2), \label{cgauge}
\end{eqnarray}
since it corresponds to the phase picked up by the fermions as a result of the
local gauge transformation.
Obviously, the Green function on the right hand side of
Eq.~(\ref{Gtransform}) is not generically invariant in this case:
\begin{eqnarray}
G ({\bf r}_{1},t_{1};{\bf r}_{2},t_{2})\rightarrow \exp [ie (\chi
({\bf r}_{1},t_{1})-\chi ({\bf r}_{2},t_{2}))]G ({\bf
r}_{1},t_{1};{\bf r}_{2},t_{2}).
\label{Ggauge}
\end{eqnarray}
However, it is possible to show that its transform in Eq.~(\ref{Gtransform})
is invariant, when one chooses\cite{bertoncini_jauho}
\begin{eqnarray}
W ({\bf k},\omega,{\bf r},t,{\bf r}_{ave},t_{ave})= \int_{-1/2}^{1/2}d\lambda
\{ t[\omega+e\varphi ({\bf r}_{ave}+\lambda {\bf r},t_{ave}+\lambda t))
\nonumber
\\
-{\bf r}\cdot({\bf k}+e{\bf A} ({\bf r}_{ave}+\lambda {\bf r},t_{ave}+\lambda t)]
\}
\label{w}
\end{eqnarray}
(for details, see Ref.~\refcite{Haug}).

In the case of a spatially homogeneous electric field in the Hamiltonian gauge
with $\varphi ({\bf r},t)=0$, which we study in this paper, this
transformation is
\begin{eqnarray}
\tilde G({\bf k},t,{\bf r}_{ave},t_{ave})\rightarrow G\left ( {\bf
k}-\frac{1}{t}\int_{-t/2}^{t/2} e{\bf A}(t_{ave}+\bar t)d\bar t,t,{\bf
r}_{ave},t_{ave}\right ), \label{eq: g_gauge}
\end{eqnarray}
because the function $W$ just involves a shift of the momentum;
note that the Green function is actually independent of ${\bf r}_{ave}$
in this case.  Hence, the gauge invariant Green function in the momentum
representation contains a shift of the momentum, which depends on both
the relative and average time coordinates.
We consider the case when a constant electric field
is turned on at time $t=0$: ${\bf A}(t)=-{\bf E}t\theta (t)$. Then the
momentum shift is
\begin{eqnarray}
{\bf k}\rightarrow {\bf k}&-&e{\bf E}\Bigr [ t_{ave}\theta(t_{ave}-|t/2|)
\nonumber\\
&+&\left (
-\frac{t_{ave}^2}{2t}+\frac{t_{ave}}{2}-\frac{t}{8}\right )
\theta(-t/2-|t_{ave}|)\nonumber\\
&+&\left (
\frac{t_{ave}^2}{2t}+\frac{t_{ave}}{2}+\frac{t}{8}\right )
\theta(t/2-|t_{ave}|)\Bigr ].
\label{shift}
\end{eqnarray}
Note that this shift does not depend on the relative time coordinate $t$ for
long times, $t_{ave}>|t/2|$. However, in general, one has to first shift
the momentum, and then Fourier transform the relative time to a
frequency. It is important that the time-dependent momentum shift takes place
for some negative average times (if the absolute value of the
relative time is large enough,
then either $t_1$ or $t_2$ is larger than 0 and hence ``sees'' the field).

The shift of the momentum becomes particularly simple for equal
time Green functions, such as those needed to calculate the
current flowing or to determine the distribution of the electrons
amongst the quantum states.  In this case, $t=0$, and the momentum
is shifted by $-e{\bf E}t_{ave}$ if $t_{ave}>0$.
Therefore, gauge invariant Green
functions can be obtained from the Hamiltonian gauge Green functions by
simply shifting the momentum by
$-e{\bf E}t_{ave}$. Note that local quantities, like the
local density of states or the local distribution function are
always gauge invariant, because they are summed over momentum, and if the
shift is the same for each momentum value, then we still sum over all the
momentum points in the Brillouin zone. In cases where the relative
time is nonzero, the transformation from the Green function in a
particular gauge to the gauge-invariant Green function must be handled
with care.  Finally, one should note that in the steady state, where
$t_{ave}\rightarrow\infty$, the momentum shift is also simple
($-e{\bf E}t_{ave}$); it turns out that the retarded and advanced Green
functions depend only on the relative time, but the lesser, greater, and
Keldysh Green functions depend on both the average and relative time
because there is an average-time-dependent shift of the momentum in
Fermi-Dirac distribution functions.  Caution must be used in trying to
directly find the steady-state Green functions, because the Dyson
equation is modified, since the momentum shift does not remove all
average time dependence in internal variables that are integrated
over in the $G_0\Sigma G$ term.

\section{Bloch electrons in infinite dimensions}
\label{Free}

The work presented in this section is based on Ref.~\refcite{Turkowski}
where the original solution for Bloch electrons in a field was
given.  There the work focused on the Hamiltonian gauge, here
we discuss the gauge-invariant formalism.


Bloch\cite{bloch} and Zener\cite{zener} originally showed that when
electrons are placed on a perfect lattice, with no scattering, the
current oscillates due to Bragg reflection of the wavevector as
it evolves to the Brillouin-zone boundary. Here we show how to analyze
this problem on the infinite-dimensional hypercubic lattice.
The noninteracting problem can be solved
exactly in the case of an arbitrary time-dependent electric field.
In particular, the noninteracting contour-ordered Green function is
(in the Hamiltonian gauge\cite{Turkowski}):
\begin{eqnarray}
G_{{\bf k}}^{c0}(t_{1},t_{2})=&~&i [f(\epsilon ({\bf
k})-\mu)-\theta_{c} (t_{1},t_{2})] \exp [i\mu (t_1-t_2)] \nonumber
\\
&\times&\exp \left[ -i\int_{t_{2}}^{t_{1}}d{\bar t}\epsilon
\left({\bf k}-e{\bf A}({\bar t}) \right) \right] , \label{Gc0}
\end{eqnarray}
where $f(\epsilon ({\bf k})-\mu)]=1/\{1+\exp[\beta(\epsilon({\bf k})-\mu)]\}$
is the Fermi-Dirac distribution (half-filling corresponds
to $\mu=0$).  The symbol $\theta_c(t_1,t_2)$ is equal to one if $t_1$ lies
after $t_2$ on the contour, and is zero otherwise.
Note that the Green function in Eq.~(\ref{Gc0}) is also used
in the system of equations (\ref{DMFT1})-(\ref{DMFT3}) to solve the
interacting problem.

When we have a constant electric field directed along the diagonal and
turned on at $t=0$, each component of the vector potential satisfies
$A(t)=-Et\theta(t)$. Then
the integral that appears in the exponent of Eq.~(\ref{Gc0}) is
\begin{eqnarray}
&~&\theta(-|t/2|-t_{ave})\epsilon({\bf k})t\label{eq: nonint_exp}\\
&+&\theta(-t/2-|t_{ave}|)\nonumber\\
&\times&\left [\frac{\epsilon({\bf k})(\sin eE(t_{ave}-t/2)
+t_{ave}+t/2)-\bar\varepsilon({\bf k})(\cos eE(t_{ave}-t/2)-1)}{eE}\right ]
\nonumber\\
&+&\theta(t/2-|t_{ave}|)\nonumber\\
&\times&\left [\frac{\epsilon({\bf k})(\sin eE(t_{ave}+t/2)
-t_{ave}+t/2)+\bar\varepsilon({\bf k})(\cos eE(t_{ave}+t/2)-1)}{eE}\right ]
\nonumber\\
&+&\theta(t_{ave}-|t/2|)\Bigr [ \epsilon({\bf k})(\sin
eE(t_{ave}+t/2)
\nonumber\\
&~&
-\sin eE(t_{ave}-t/2))+\bar\varepsilon({\bf k})(\cos eE(t_{ave}+t/2)-
\cos eE(t_{ave}-t/2))\Bigr ] \Bigr / eE\nonumber
\end{eqnarray}
when expressed in terms of the Wigner coordinates.

To get the gauge-invariant Green function, we now shift the momentum
as shown in Eq.~(\ref{shift}); note that the shift is done both for
the momentum in the exponent, and for the momentum in the
Fermi-Dirac distribution. Two of the four cases for the exponent in
Eq.~(\ref{eq: nonint_exp}) are easy to work out for the
gauge-invariant Green functions.  The first is the
$\theta(-|t/2|-t_{ave})$ term which remains unchanged and the second
is the $\theta(-t/2-|t_{ave}|)$ term, which becomes $2\epsilon({\bf
k})\sin (eEt/2)$.  Note that both these exponents are independent of
the average time.  The average time enters for the other two terms,
and in the argument of the Fermi-Dirac distribution.

The retarded $G_{{\bf k}}^{R0}(t_{1},t_{2})$  and lesser $G_{{\bf
k}}^{<0}(t_{1},t_{2})$ Green functions can be obtained from
Eq.~(\ref{Gc0}), by replacing the prefactor $[f(\epsilon
({\bf k})-\mu)-\theta_{c} (t_{1},t_{2})]$ by $-i\theta
(t_{1}-t_{2})$ and $f(\epsilon ({\bf k})-\mu)$, correspondingly.
One needs to shift the momentum accordingly to get the gauge-invariant
retarded and lesser Green functions. Note that at long times the
gauge-invariant retarded Green function depends only on relative time.


Since the electrical current is found from the time derivative of
the polarization operator, the current operator is determined by taking
the commutator of the Hamiltonian (in a particular gauge) with the
polarization operator.  The result, for the
$\alpha$th component of the current-density operator is
\begin{equation}
{\bf j}_\alpha(t_{ave})=e\sum_{\bf k} \frac{\partial \epsilon ( {\bf
k}- e{\bf A}(t_{ave}))}{\partial k_{\alpha}} c^\dagger_{\bf k}(t_{ave})c_{\bf
k}(t_{ave}) \label{j},
\end{equation}
where we have emphasized that the operator is evaluated with a
vanishing relative time ($t=0$).
We want the expectation value of the current operator, which is found
by taking expectation value of the expression in Eq.~(\ref{j}) and
noting that each component gives the same result for a field pointing
along the diagonal. The
expectation value of the product $c^\dagger_{\bf k}(t_{ave})c_{\bf k}(t_{ave})$
can be replaced by the lesser Green function $G^<_{\bf k}(t_{ave},0)$.
So we have
\begin{equation}
j(t_{ave}) =ie \sum_{{\bf k}} {\bar \varepsilon } \left(
{\bf k}-e{\bf A}(t_{ave})\right) G_{{\bf k}}^{<}(t_{ave},0)
=ie \sum_{{\bf k}} {\bar \varepsilon } \left(
{\bf k}\right) \tilde G_{{\bf k}}^{<}(t_{ave},0), \label{jalpha2}
\end{equation}
where the second equality comes from the transformation to the
gauge-invariant Green function.
The summation over momentum can be converted to a double integral over
the joint density of states in Eq.~(\ref{rho2}).
Substituting in the expression for the lesser
Green function, yields
\begin{eqnarray}
j(t_{ave})=\frac{et^*}{4\sqrt{d}\pi} \sin (eEt_{ave})\int d\epsilon \frac{d
f(\epsilon-\mu)}{d\epsilon}\rho(\epsilon),
\end{eqnarray}
where the single-particle density of states is
\begin{eqnarray}
\rho (\epsilon ) =\int d{\bar\varepsilon }\rho_{2}(\epsilon , {\bar
\varepsilon}) =\frac{1}{\sqrt{\pi}}\exp\left[ -\epsilon^{2} \right]
. \label{rho}
\end{eqnarray}
The current is a periodic function of
time, even though the field is time-independent; this effect is
called a Bloch oscillation\cite{ashcroft_mermin_1976}. The
period of the oscillation is equal to $2\pi /eE$. In order to see
this oscillation in real solids, one needs to prepare a system where
the scattering time is longer than the period of the Bloch
oscillations. In solids, the scattering
time is much shorter than the oscillation period, so this effect is not
observed. However, Bloch oscillations are seen in
semiconductor superlattices, where the period of oscillations is much
shorter due to the larger lattice spacing.
As we show in the
following Sections, the effects of strong electron-electron
correlations modify the Bloch oscillations significantly, but
the driven oscillations in large fields survive for a surprisingly long time.


Now we discuss the time-dependence of the density of
states (DOS) for noninteracting electrons in a
constant electric field. The DOS is found
by using the Wigner time coordinates in Eq.~(\ref{tWigner}), and making a
Fourier transformation of the corresponding Green functions with respect
to the relative time coordinate. In particular, the local DOS is
\begin{equation}
A(t_{ave},\omega)=-\frac{1}{\pi}{\rm Im} \int_{0}^{\infty} d t
e^{i\omega t} G^R_{loc}(t_{ave},t). \label{At}
\end{equation}
Since this is a local quantity, summed over all momenta, it is automatically
gauge-invariant.  The local retarded Green function can be obtained from
Eq.~(\ref{Gc0}). It is possible to show\cite{Turkowski} that the
steady state ($t_{ave}\rightarrow\infty$), for
the case of a constant electric field turned on at $t=0$, has a retarded
Green function which satisfies
\begin{eqnarray}
G^R_{loc}(t_{ave}\rightarrow\infty,t)=
-i\theta(t)\exp\left [ \frac{1}{2e^2E^2}\left \{
\cos(eEt)-1\right \}\right ].
\end{eqnarray}
Substitution of this expression into Eq.~(\ref{At}), and evaluating the
Fourier transform with respect to the relative time, yields the steady
state DOS, which consists of a set
of delta-functions with different amplitudes (called the Wannier-Stark
ladder\cite{wannier_1962}). The distance between
the delta-function peaks is equal to $eE$.
The weight of these peaks is\cite{Turkowski}:
\begin{eqnarray}
w_N=\frac{2}{e^2E^2}\int_0^{2\pi}du\cos(Nu)
\exp(\frac{t^{*2}}{2e^2E^2} [\cos u-1]),
\end{eqnarray}
for the $N$th Bloch frequency, $\omega_N=eEN$.
It takes an infinite amount of time for the delta functions to develop.

\begin{figure}
\centerline{\psfig{file=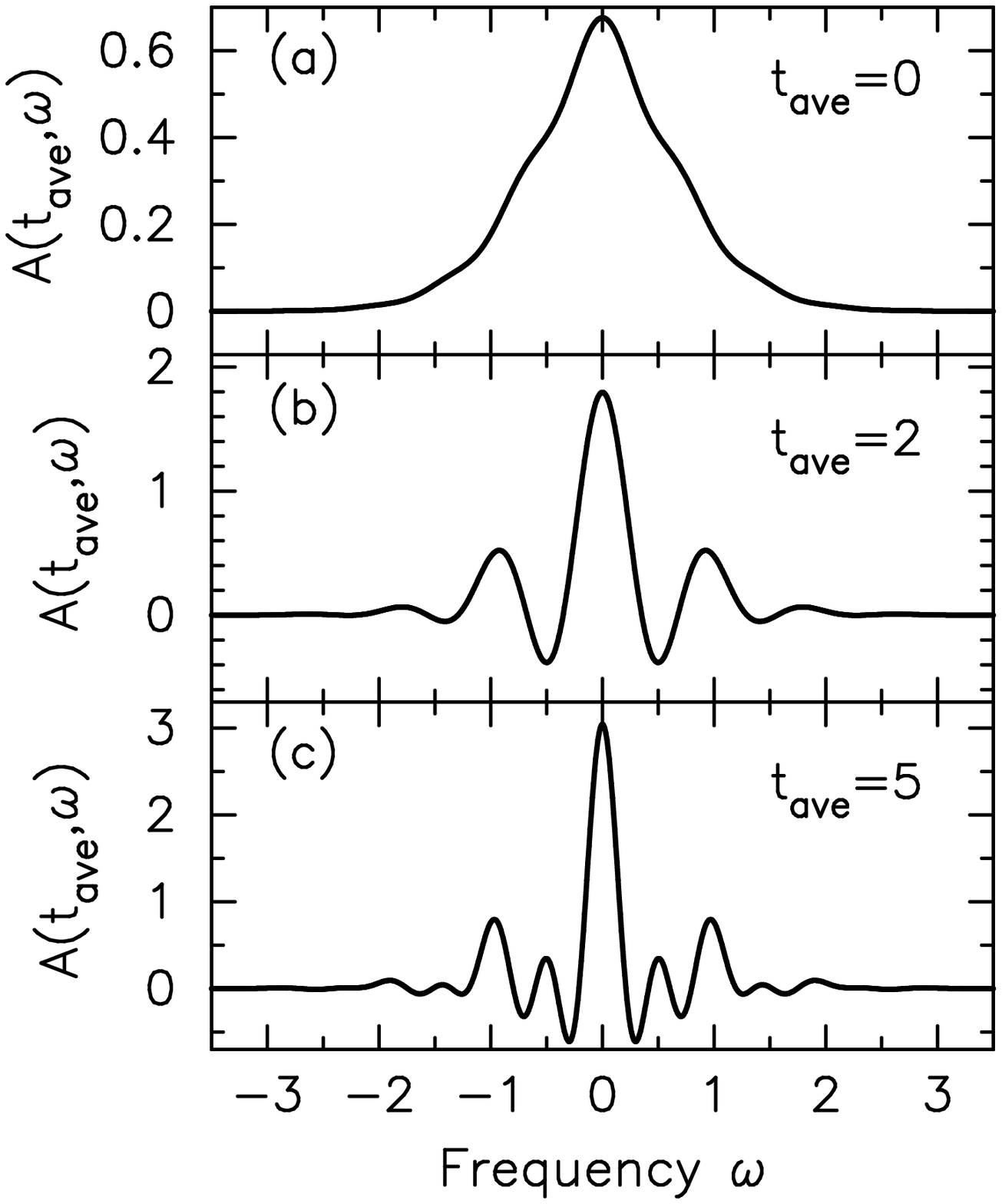,width=5.8cm}
\psfig{file=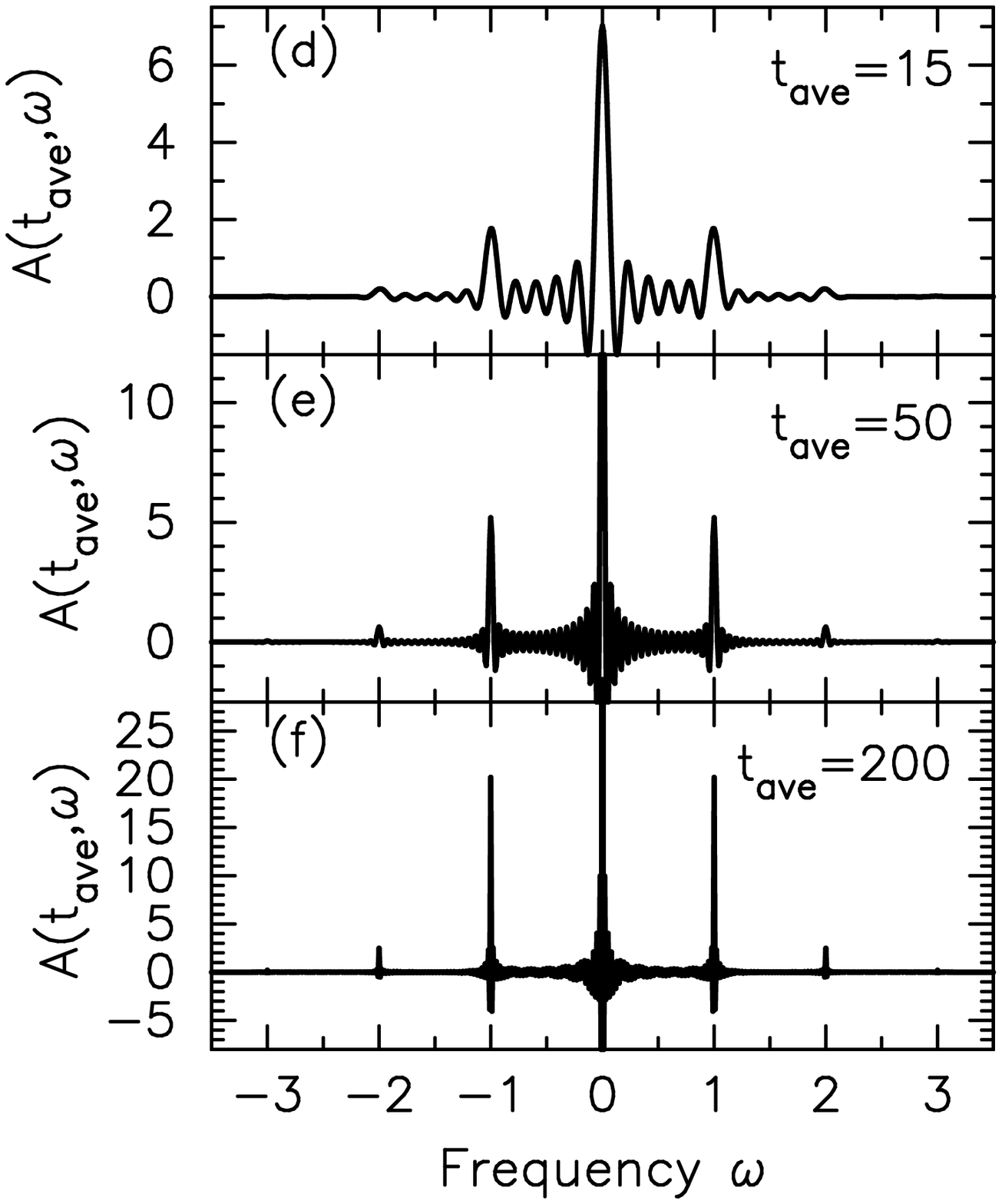,width=5.8cm}} \caption{Density of
states for noninteracting electrons with $eE=1$ at different
values of time $t_{ave}$ (be aware that the vertical scale changes from
plot to plot). Note how the build up of the delta function
at the Bloch frequencies is slow.} \label{fig: dos}
\end{figure}

In Fig.~\ref{fig: dos}, we show how the DOS evolves from the time
the field is turned on, at $t_{ave}=0$, to a large time.
The DOS remains Gaussian for
$t_{ave}<-2$ and then develops large oscillations as $t_{ave}$
increases. Though the DOS oscillates and
acquires negative values in the transient regime, it is possible to show
(numerically) that its first three moments always satisfy the relevant
sum rules.

\section{Exact solution}
\label{Exact}

In this Section we present the results for the interacting
case\cite{Nashville,Denver,dmft_fk},
where we vary the Coulomb repulsion through the metal-insulator
transition that occurs at $U_{c}=\sqrt{2}$.
The problem is solved by numerically solving the DMFT loop
in Eqs.~(\ref{DMFT1})-(\ref{DMFT3}).

\begin{figure}
\centerline{\psfig{file=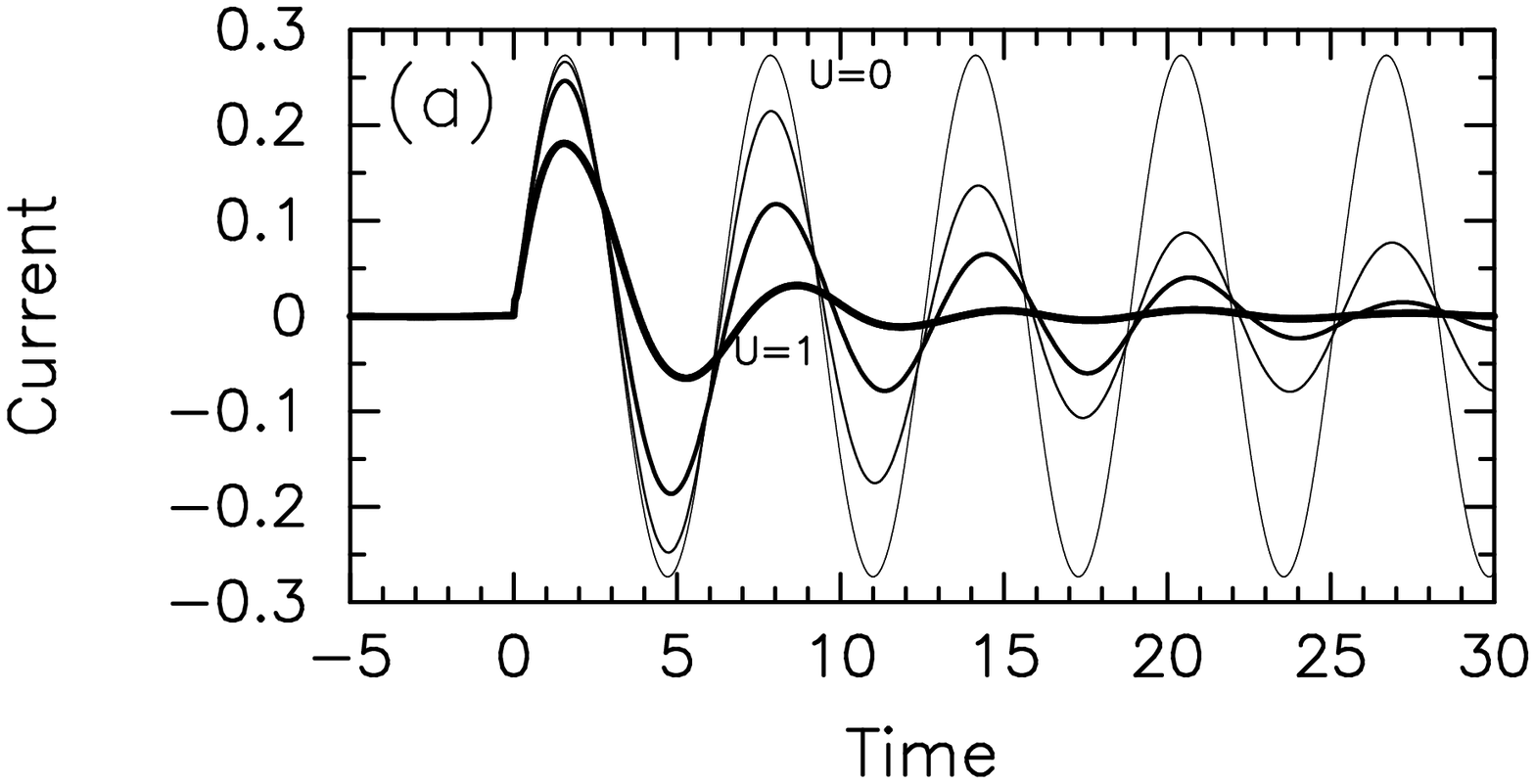,width=5.8cm}
\psfig{file=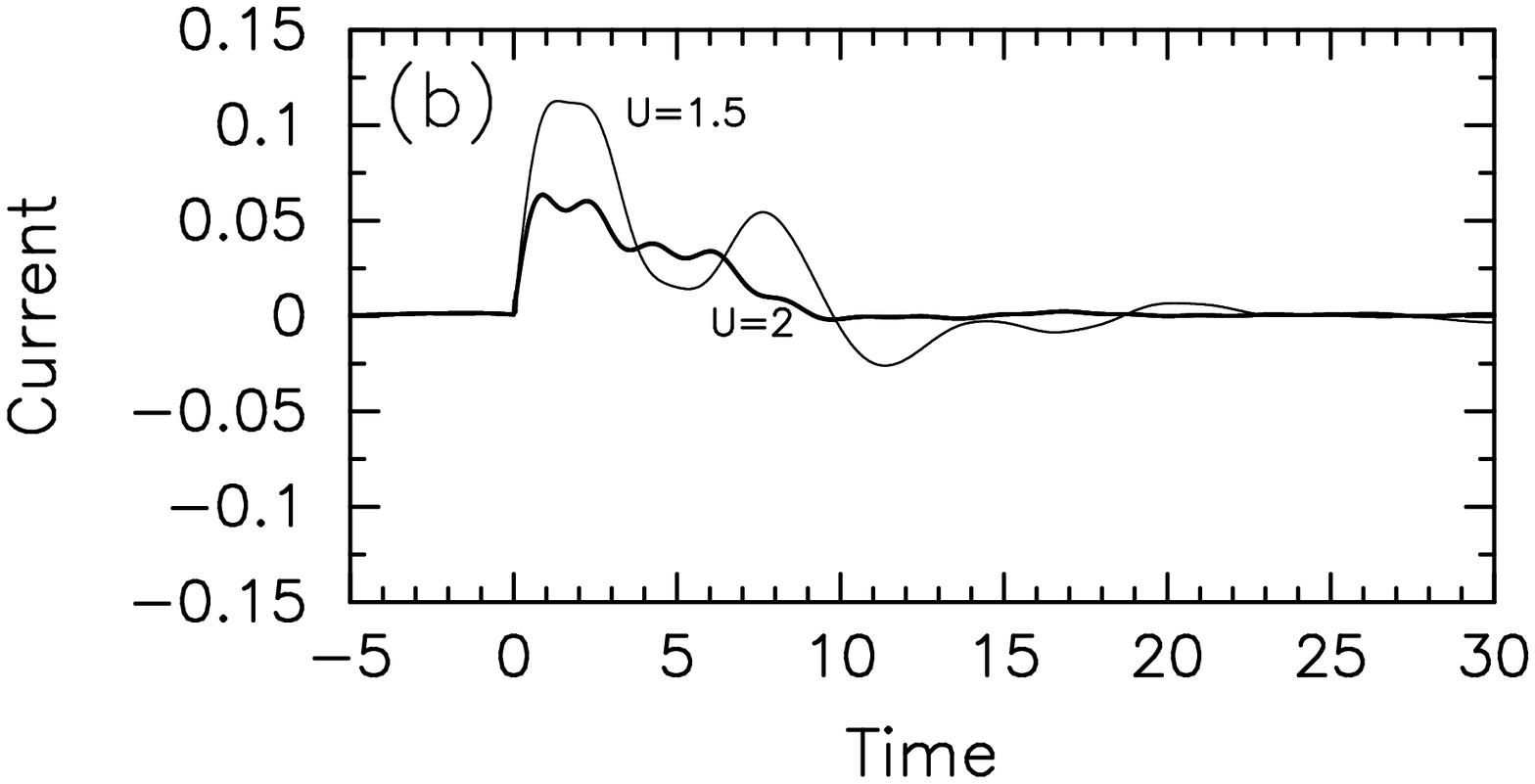,width=5.8cm}} \caption{Electric current for
different values of $U$ with $\beta=10$: (a) metals ($U=0$,
$U=0.25$, $U=0.5$, and $U=1$) and (b) insulators ($U=1.5$ and
$U=2$).} \label{JU}
\end{figure}

Once the Green functions and self-energy have been found by
self-consistently solving the DMFT equations, we can extract the
momentum-dependent lesser Green function and use
Eq.~(\ref{jalpha2}) to find the current.  In these calculations,
we used the Green functions in a particular gauge, but one could easily
shift to the gauge-invariant Green functions if desired.
When the field is small, and the correlations are small, we see a damping of
the Bloch oscillations, as expected.   This is shown in the left panel
of Fig.~\ref{JU}.  One can see that the Bloch oscillations maintain
their periodicity, but are damped as the scattering increases.  As
we start to approach the metal-insulator transition at $U\approx 1.414$,
one can see the character of the oscillations start to change.  As
we move into the insulating phase, as shown in the right panel, the
character of the oscillations changes completely, and we no longer
see the regular Bloch structure.  The oscillations seem to survive
to much longer times than would be expected from a Boltzmann equation
type of analysis.  It remains unclear whether the steady state has
some residual oscillations, or it goes to a constant value as predicted
by semiclassical ideas.

\begin{figure}
\centerline{\psfig{file=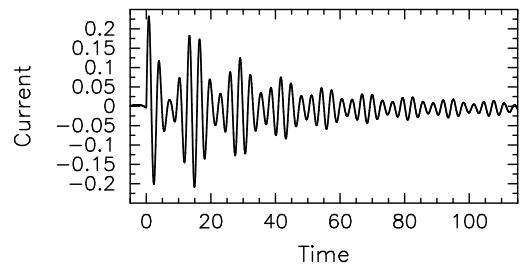,width=7.0cm}} \caption{Time-dependence
of the current for $U=0.5$, $E=2.0$, and $\beta =10$. Note how the
current has beats in its time dependence and that the decay of
the current is rather slow.} \label{beatings}
\end{figure}

Even more surprising is the fact that when the field
is large, the current displays two anomalous features: (i) first, its
decay is much slower than expected from a semiclassical approach, where
the relaxation time is inversely proportional to the imaginary
part of the self-energy at the chemical potential, which is proportional
to $1/U^2$ and (ii) the current develops beats with a beat frequency
proportional to $1/U$. An example of this behavior is shown in
Fig.~\ref{beatings}. These beats are always present in the metallic
phases (for large $U$), but disappear once one moves into the insulator.

The time-dependence of the density of states can be calculated from
Eq.~(\ref{At}). Here, we present some results for the case when the
system is initially in the metallic phase, Fig.~\ref{fig:
dos_u=0.5_e=1}.  What we find is that for small fields, the delta
function peaks of the Wannier-Stark ladder get broadened, but the structure is
still readily apparent.  But as we increase the field strength, the
behavior qualitatively changes, and in the long-time limit, the
system evolves into a peaked structure, where the peaks are maximal near
the edges of minibands, which are spaced apart in size by $U$, and the
DOS has a local minimum in the center, where the Wannier-Stark peak
used to appear. This is also behavior that is quite surprising.

\begin{figure}
\centerline{\psfig{file=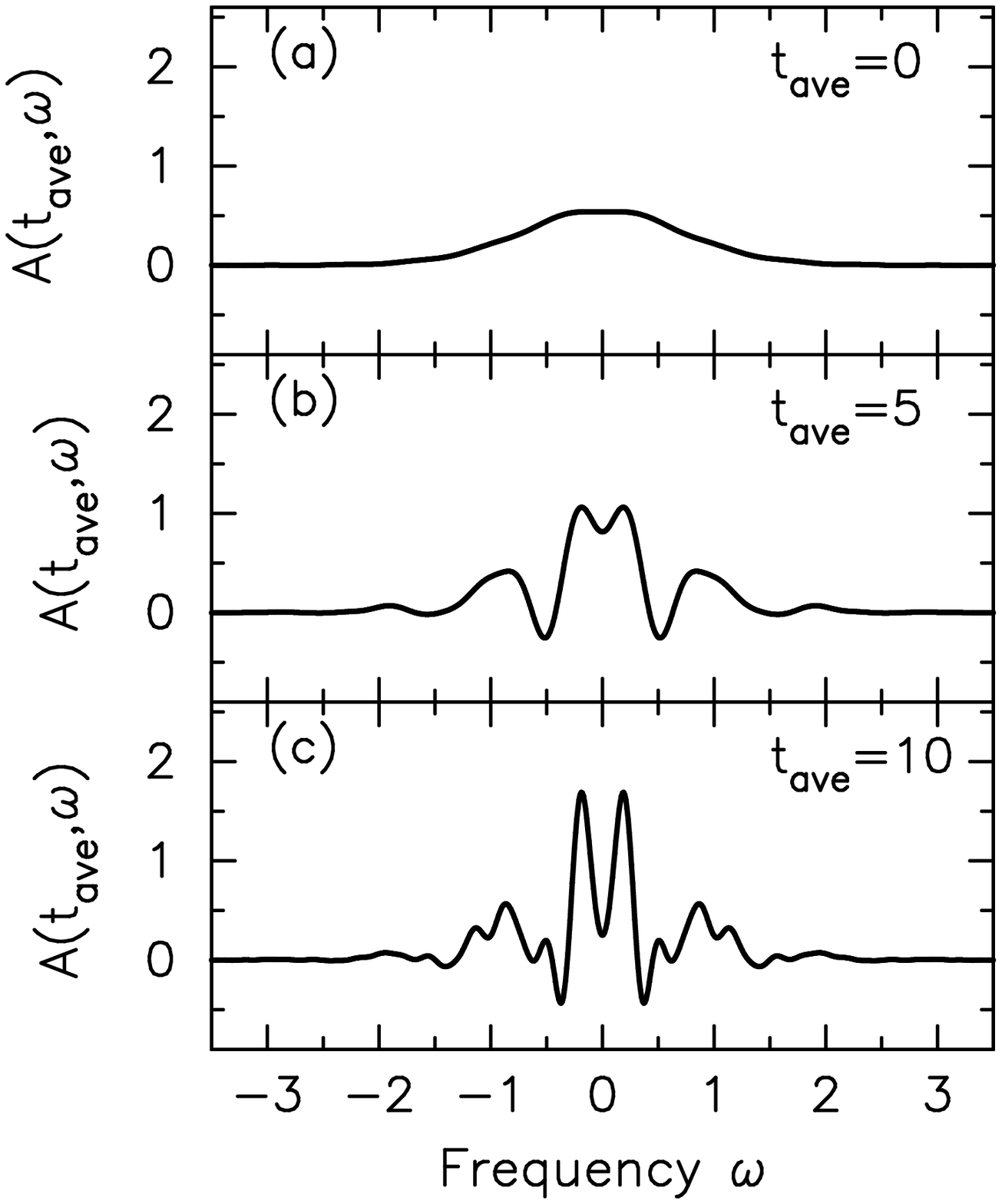,width=5.8cm}
\psfig{file=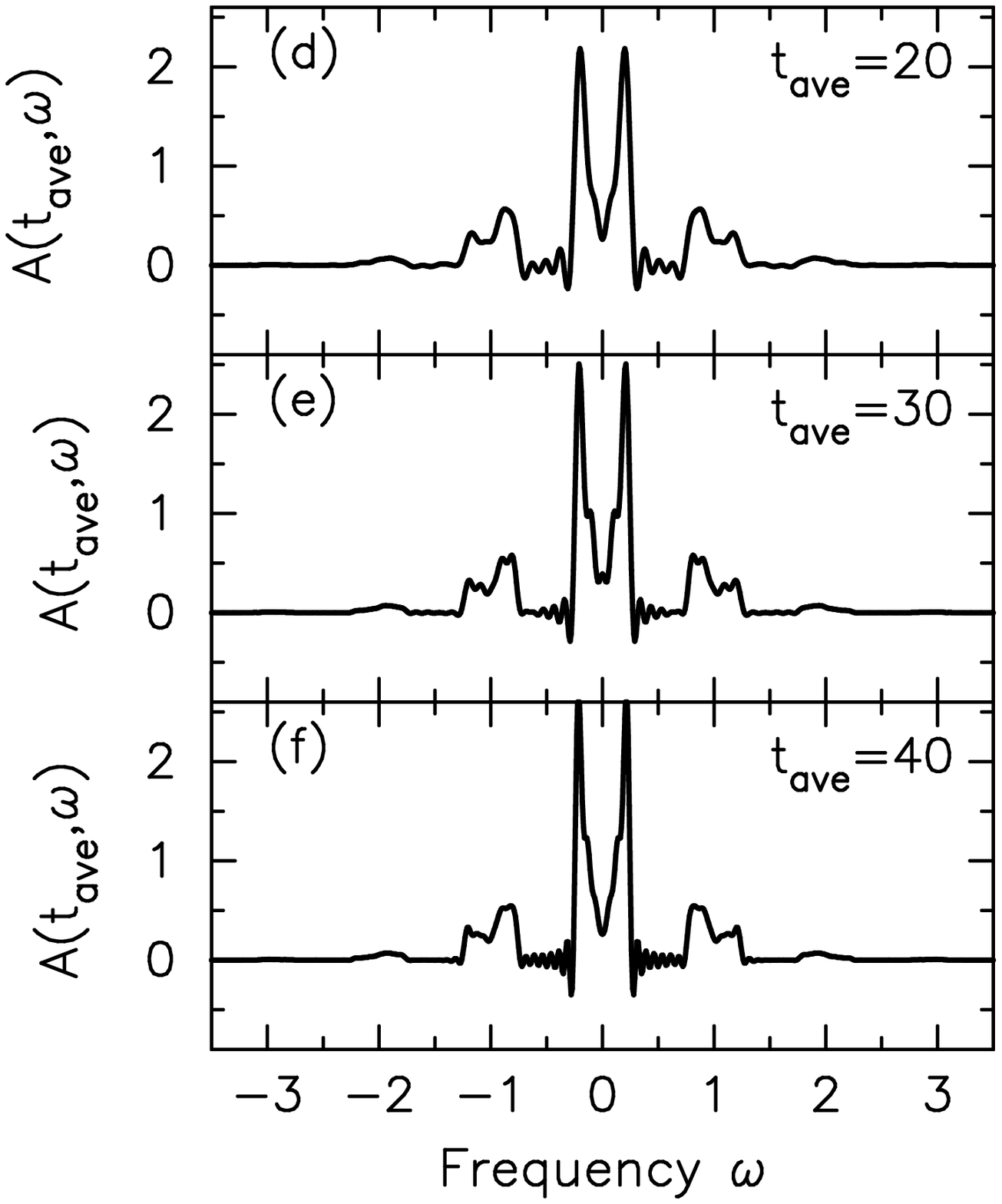,width=5.8cm}} \caption{Density of states for
different average times from $t_{ave}=0$ to $t_{ave}=40$ for
$U=0.5$, $E=1$ and $\beta=10$. Note how the DOS develops split
peaks, separated by $U=0.5$ around the Bloch frequencies (integers
here)} \label{fig: dos_u=0.5_e=1}
\end{figure}

As the scattering increases, the DOS approaches the steady state value
relatively quickly.  This illustrates the dichotomy between the
average time, which is important for determining the current, and
the relative time, which is important for determining the DOS.  The
decay is rapid as a function of relative time, but is much slower
as a function of average time.

\section{Perturbation theory}
\label{Perturbationtheory}

A perturbative analysis can be performed directly on the
lattice\cite{PT}.  In this case, we do not need any DMFT loop,
and we can restrict the contour to be solely on the real axis.
As described above, the perturbation theory is similar for the
equilibrium and nonequilibrium cases, with the only significant changes
being that one needs to calculate with time-ordered objects along the contour
and one needs to use noninteracting Green functions in the field.
A strictly truncated expansion for the self-energy to second order
in $U$ is equal to the usual Hartree-Fock term (which vanishes at half
filling) plus a second-order term which satisfies
\begin{equation}
\Sigma^{c(2)}(t_1,t_2)=U^2 w_1(1-w_1) G^c_{loc}(t_1,t_2);
\label{eq: self_pert}
\end{equation}
one can determine the retarded and lesser self-energies from this in
a straightforward fashion.  It turns out that this truncated
perturbation theory is most accurate at short times---in essence, the
perturbation series expansion is an expansion in a power series in
time away from the time the field was turned on.  In more conventional
perturbation series in terms of frequency-dependent Green functions,
the perturbation series is most accurate at high frequencies, and least accurate
at low frequencies.  After performing a Fourier transform, one can
immediately see that this is equivalent to the perturbation theory being
most accurate at short times and breaking down at long times.  Indeed,
we find that this perturbative treatment cannot reproduce the steady-state
behavior at long times.

As a benchmark of our calculation, we compare the equilibrium self-energy
found from a numerical solution of the DMFT equations to the perturbative result
for small $U$.  We find quite good agreement in the small-$U$ range, and
the equilibrium case appears to be fairly accurate up to $U\approx 0.5$.  For
larger $U$ values the perturbation theory breaks down---it is not capable
of properly describing the Mott-insulating phase.

Next we analyze the time-dependence of the electric
current calculated by second order perturbation theory in the
case when a constant electric field is turned on at time $t=0$.
Before presenting the
second-order perturbative solution for the current, we briefly review
the corresponding results from a semiclassical Boltzmann
equation approximation. As was mentioned above,
these results are qualitatively different from the exact solution.

In the Boltzmann equation approach, one introduces a nonequilibrium
quasiparticle distribution function $f^{\rm non}({\bf
k},t)=-iG_{{\bf k}}^{<}(t,t)$, which satisfies the following
phenomenological equation:
\begin{equation}
\frac{\partial f^{\rm non}({\bf k},t)}{\partial t} +e{\bf
E}(t)\cdot\nabla_{\bf k}f^{\rm non}({\bf k},t)
=-\frac{1}{\tau}[f^{\rm non}({\bf k},t)-f({\bf k})],
\label{Boltzmann}
\end{equation}
with the boundary condition:
\begin{eqnarray}
f^{\rm non}({\bf k},t=0)=f({\bf k})=\frac{1}{\exp[\beta
(\epsilon({\bf k}) -\mu)]+1} . \label{Boltzmannbc}
\end{eqnarray}

This equation can be solved exactly (see, for example
Ref.~\refcite{PT}). Substitution of the expression for the
distribution function instead of $-iG^{<}$ into Eq.~(\ref{jalpha2})
allows one to calculate the semiclassical current. This
semiclassical current approaches a steady state as time goes to
infinity. In particular, in the infinite-dimensional limit one
obtains:
\begin{eqnarray}
j(t)&=&-\frac{e}{\sqrt{d}}\frac{eE\tau}{1+e^{2}E^{2}\tau^{2}} \int
d\epsilon \rho (\epsilon )\epsilon f(\epsilon) \nonumber\\
&\times&\left[ 1-\left(\cos (eEt)-eE\tau\sin (eEt) \right)
e^{-t/\tau} \right] . \label{jBoltzmannsol}
\end{eqnarray}
Therefore, the current is a strongly oscillating function of time
for $t\ll\tau$, and it approaches the steady-state value
\begin{equation}
j^{\rm steady}=\frac{eE\tau}{1+e^{2}E^{2}\tau^{2}}j_{0},
\label{jBoltzmannsteady}
\end{equation}
where
\begin{equation}
j_{0}=-\frac{e}{\sqrt{d}}\int d\epsilon \rho (\epsilon
)\epsilon f(\epsilon ), \label{j0}
\end{equation}
as $t/\tau\rightarrow\infty$. The steady-state current amplitude is
proportional to $E$ in the case of a weak field (the linear-response
regime), and then becomes proportional to $1/E$ at
$eE\tau\rightarrow\infty$. The amplitude of the current goes to zero
in this nonlinear regime with the field amplitude increasing. One
would naively expect that the second-order perturbation theory would
give similar results in the case of a weak Coulomb repulsion, since
one can extract an effective scattering time for the equilibrium
limit of the Falicov-Kimball model  with small $U$:
$\tau=1/(\pi^{2}U^{2})$ \cite{Kiel}. However, it will be shown
below, that the behavior of the current calculated in second-order
perturbation theory is rather different from the Boltzmann equation
case, and closer to the exact numerical result at short times.

The electric current in the second-order perturbation theory can be
calculated by substituting the expression for the second-order
lesser Green function into Eq.~(\ref{jalpha2}). In this case
\begin{eqnarray}
j(t) =\frac{ie}{\sqrt{d}} \int d\epsilon\int d{\bar
\varepsilon}\rho_{2}(\epsilon ,{\bar\varepsilon})\left[{\bar \varepsilon}\cos
\left( eA_{\alpha}(t)\right) -\epsilon\sin \left(
eA_{\alpha}(t)\right) \right] G_{\epsilon ,{\bar
\varepsilon}}^{<}(t,t).\nonumber \\
\label{j2}
\end{eqnarray}

It is difficult to find exact analytical expressions for the
current, except for some limiting cases. Of course,
in the limit $U=0$ we recover the free electron case result:
\begin{eqnarray}
j(t)=j_{0}\sin \left( eEt\right) , \label{jU0}
\end{eqnarray}
where the amplitude $j_0$ of the Bloch oscillations
is given by Eq.~(\ref{j0}). Therefore, the general
expression for the time-dependence of the electric current in a
strictly truncated second-order perturbation theory expansion can be
written as:
\begin{eqnarray}
j(t)=j_{0}\sin \left( eEt\right)+U^{2}j_{2}(t). \label{jU}
\end{eqnarray}
The electric current is a superposition of an oscillating part and
some other piece proportional to $U^{2}$. Obviously this cannot
produce a constant steady-state current for all small $U$, because
the function $j_2$ is independent of $U$.  This is a clear indication
that the perturbation theory will hold only for short times.

\begin{figure}
\centerline{\psfig{file=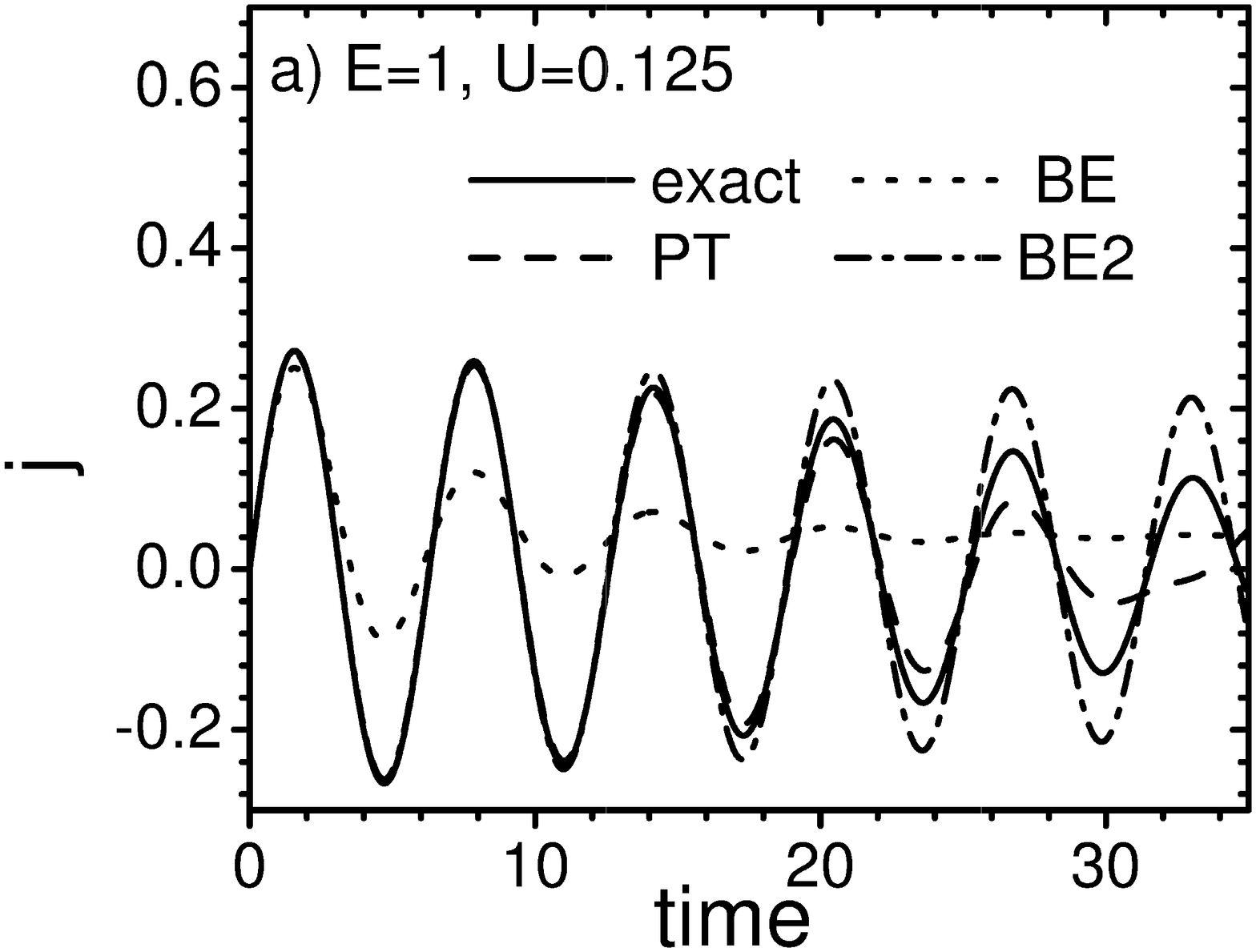,height=3.9cm, angle=270}
\psfig{file=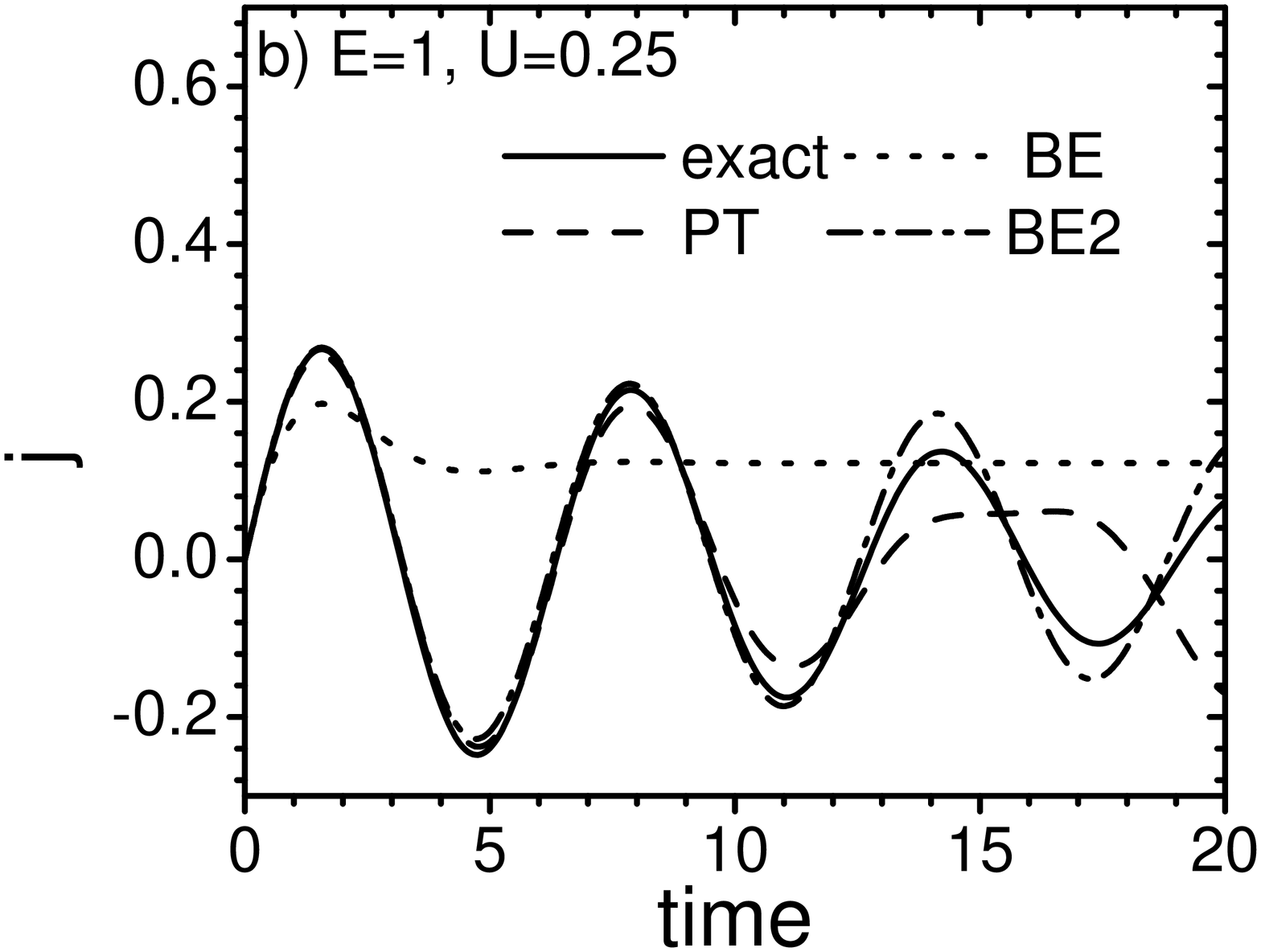,height=3.9cm, angle=270}
\psfig{file=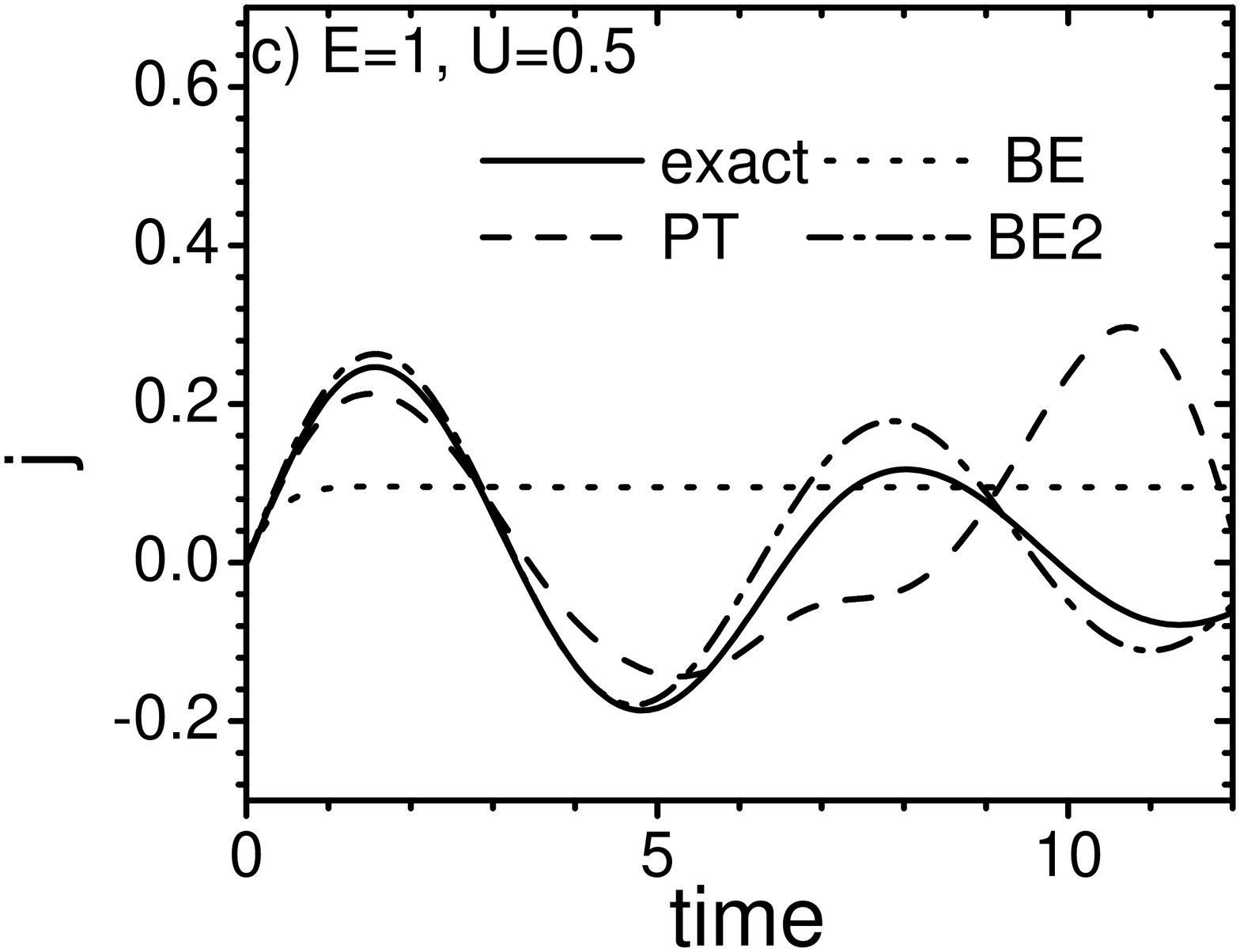,height=3.9cm, angle=270}} \caption{Perturbative
expansion for the electric current as a function of time for
$E=1.0$, $\beta =10$ and different values of $U$ (dashed lines). The
solid and dotted lines correspond to the exact DMFT solution and the
Boltzmann equation (BE) solution, respectively. The dash-dotted
lines (BE2) are the Boltzmann equation result with a
phenomenological relaxation time $\tau =\alpha/(\pi^{2}U^{2})$,
while $\alpha=20$ in Figs.~a) and b), and $\alpha =36.5$ in
Fig.~c).} \label{fig:current}
\end{figure}

Numerical results for the time-dependence of the electric current
calculated from Eq.~(\ref{j2}) at $eE=1$ and different values of $U$
are presented in Fig.~\ref{fig:current} (dashed lines).
Note how the current oscillates for all times within our finite
time window.
We also show the corresponding Boltzmann equation solution and
the exact solution. We use two different values for the Boltzmann
equation---one fixes the relaxation time to the prediction from the
equilibrium solution, while the other adjusts the relaxation time to obtain the
best fit. Comparison of the perturbation theory
and the Boltzmann equation solution shows that they are close at
short times, but at longer times the PT current remains
oscillating, while the Boltzmann equation solution
approaches a steady state. Moreover, at times longer than $\sim
2/U$ the perturbation theory breaks down showing an oscillating
current with increasing amplitude. At times shorter than $2/U$ the
perturbation theory solution is close to the exact result, displaying
an oscillating current with decreasing amplitude. It is also
possible to fit the Boltzmann equation results to the exact
and PT solution at short times if one chooses the relaxation time
$\tau=\alpha /(\pi^{2}U^{2})$, where $\alpha \sim 20-30$, which is
much larger than $\alpha =1$ in the case of the second order
perturbation theory\cite{Kiel}. These results clearly show that the
semiclassical approach, with one effective time variable that is damped on the
timescale of the relaxation time is not sufficient to describe the
behavior in the quantum case.

The perturbation theory calculations at different values of the
electric field give results similar to the results presented in Fig.
\ref{fig:current}. Numerical analysis shows that the agreement
between the perturbation theory results and the exact results is
better when $eE$ is larger than $U$. In fact, it is possible to find
an analytical expression for the current in the case of large
electric fields:
\begin{eqnarray}
j(t)\simeq -\frac{e}{\sqrt{d}}\int d\epsilon \rho (\epsilon )
\epsilon f(\epsilon )\left[1-U^{2}B(\beta
)-\frac{U^{2}}{4}t^{2} \right] \sin (eEt), \label{jstrongfields}
\end{eqnarray}
where $ B(\beta )$ is a positive decreasing function of temperature\cite{PT}:
$0.25<B(\beta )<0.5$.

\section{Conclusions}
\label{Conclusions}

To conclude, we have presented some results on the nonequilibrium
properties of the Falicov-Kimball model of strongly correlated
electrons in the limit of infinite dimensions. Despite the
simplicity of the model, the solutions show that strong
electron-electron correlations result in nontrivial behavior. The
dynamical mean-field theory approximation is believed to be a
precise method to solve strongly correlated problems in three
dimensions. Therefore, we believe that some of the results, like the
long-time Bloch oscillations of the current, beats in the current
for strong fields and the splitting of the Wanier-Stark peaks could
be observed in bulk systems with dominant electron-electron
scattering in the presence of a strong electric field. Such a field
can be present, in particular, in nanostructures, where a moderate
external electric potential can produce strong (uniform) electric
fields due to the small size of the systems. One might also be able
to observe this behavior in mixtures of heavy and light atoms
trapped in optical lattices. In addition, we demonstrated that the
perturbation theory solution cannot be used to study the long-time
behavior of the system. It would be interesting to generalize these
results to more complicated models and to lower dimensions, where we
expect qualitatively similar behavior.

\section*{Acknowledgments}
We thank Antti-Pekka Jauho, Alexander Joura, Joseph Serene and Veljko
Zlati\' c for valuable discussions. We would like to acknowledge
support by the National Science Foundation under grant number
DMR-0210717 and by the Office of Naval Research under grant number
N00014-05-1-0078. Supercomputer time was provided by the DOD HPCMO
at the ASC and ERDC centers (including a 2006 CAP project)
and by a National Leadership Computing System grant from NASA.

\end{document}